\newcommand{\CLASSINPUTtoptextmargin}{19.1mm}
\newcommand{\CLASSINPUTbottomtextmargin}{19.1mm}
\newcommand{\CLASSINPUToutersidemargin}{19.1mm}
\newcommand{\CLASSINPUTinnersidemargin}{19.1mm}
\documentclass[final,twocolumn,letterpaper,conference]{IEEEtran} 
\title{\vspace{6.3mm}Generalized Geometric Programming\\for Rate Allocation in Consensus}
\IEEEoverridecommandlockouts

\author{
    Ryan Pilgrim,$^{\dagger}$ Junan Zhu,$^{\dagger}$ Dror Baron$,^{\dagger}$ and
    Waheed U. Bajwa$^{\ddagger}$
\thanks{RP, JZ, and DB were supported by the National Science
    Foundation (NSF) under
grants CCF-1217749 and ECCS-1611112. WB was supported by the NSF 
under grant CCF-1453073. RP is currently with Vadum, Inc., Raleigh, NC,
and JZ is currently with J.P.Morgan Chase, New York, NY.}\\
$^\dagger$Department of Electrical and Computer Engineering, NC State University,
Raleigh, NC \\
$^{\ddagger}$Department of Electrical and Computer Engineering, Rutgers
University, Piscataway, NJ \\
Email: \{rzpilgri, jzhu9, barondror\}@ncsu.edu, waheed.bajwa@rutgers.edu
\vspace{-3pt}
}

\usepackage{graphicx}
\usepackage{soul}
\usepackage{epstopdf}
\usepackage{floatrow}

\usepackage[utf8]{inputenc}
\usepackage{csquotes}
\usepackage{bm}
\usepackage{url}
\makeatletter
\def\tok@scan#1{%
	\ifx#1\relax
	\let\tok@next\relax
	\else
	\edef\my@list{\my@list#1}%
	\let\tok@next\tok@scan
	\fi
	\tok@next
}
\newcommand{\@strip}[2]{%
	\def\my@list{}\tok@scan#2\relax\let#1\my@list}
\newcommand{\Cite}[1]{\@strip\@args{#1}\cite\@args}
\makeatother

\usepackage[format=plain,justification=justified,singlelinecheck=false,font=small,labelfont=bf,labelsep=space]{caption} 
\usepackage{epstopdf}

\usepackage{calc}
\usepackage{citesort}
\usepackage{mathtools}
\usepackage{amsmath,amssymb,amsfonts,cleveref,autonum}
\usepackage{mathabx}
\usepackage{color}

\crefrangeformat{equation}{(#3#1#4)--(#5#2#6)}
\crefmultiformat{equation}{(#2#1#3)}%
{ and~(#2#1#3)}{, (#2#1#3)}{ and~(#2#1#3)}

\newcommand{\norm}[1]{\lVert#1\rVert}

\newcommand{\reals}{{\mathbb R}}

\newcommand{\tp}{^{\top}}

\newcommand{\fourspace}{\,\,}
\newcommand{\tindex}[1]{(#1)}
\newcommand{\vindex}[1]{_{#1}}
\newcommand{\eightspace}{\fourspace\fourspace}

\newcommand{\A}{{\mathbf{A}}}
\newcommand{\z}[1]{{{\bf z}\tindex{#1}}}

\newcommand{\q}[1]{{Q(#1)}}
\newcommand{\W}[1]{{{\bf W}^{#1}}}

\newcommand{\I}{{\bf I}}

\newcommand{\MSE}[1]{{\mathrm{MSE}}\tindex{#1}}

\newcommand{\E}[1]{{\mathbb{E}\left[#1\right]}}
\newcommand{\tr}[1]{{\mathrm{tr}}\left(#1\right)}

\newcommand{\ones}{{\mathbf{1}}}
\newcommand{\avg}{{m^{-1}{\ones \ones \tp}}}

\renewcommand{\th}{{\text{th}}}
\newcommand{\meanz}[1]{{{\boldsymbol{\mu}}}}

\DeclareMathOperator{\diag}{diag}
\DeclareMathOperator*{\argmin}{arg\,min}
\newcommand{\diff}[1]{{{{\mathbf{e}}\tindex{#1}}}}

\newcommand{\M}{{\left( \I - \avg \right)}}

\newcommand{\zmarg}[2]{{z\vindex{#1}\tindex{#2}}}

\newcommand{\margvar}[2]{{\nu_{#1}\tindex{#2}}}

\newcommand{\ze}[1]{{\boldsymbol{\zeta}\tindex{#1}}}

\newcommand{\Om}{{\boldsymbol{\Omega}}}

\newcommand{\WsubI}{{\left( \W{} - \I \right)}}

\newcommand{\Dv}{\mathbf{d}}

\newcommand{\x}{{\mathbf x}}
\newcommand{\Dmarg}[2]{{D_{#1}\tindex{#2}}}

\newcommand{\Rhat}{{\mathbf{r}}}

\newcommand{\etal}{{\em{et al.}}}

\newcommand{\mean}[2]{\boldsymbol{\mu}_{#1}\tindex{#2}}
\newcommand{\meant}[2]{\boldsymbol{\mu}_{#1}\tp\tindex{#2}}

\newcommand{\Ragg}{R_{\text{agg}}}

\newcommand{\qes}[1]{\boldsymbol{\epsilon}(#1)}
\newcommand{\qesu}{\boldsymbol{\epsilon}}

\newcommand{\Sig}[2]{{\boldsymbol{\Sigma}_{#1}\tindex{#2}}}

\newcommand{\seeqs}{eq:se-data-mean-simple,eq:se-error-mean-simple,eq:se-data-simple,eq:se-error-simple}

\begin{document}
\maketitle
\begin{abstract}
Distributed averaging, or distributed average 
consensus, is a common method for computing 
the sample mean of the data dispersed among the nodes of a network in a 
decentralized manner. By iteratively exchanging messages 
with neighbors, the nodes of the network can converge to an agreement 
on the sample mean of their initial states. In real-world scenarios, these 
messages are subject to bandwidth and power constraints, which motivates the 
design of a lossy compression strategy. Few prior works consider
the rate allocation problem from the perspective of constrained optimization,
which provides a principled method for the design of lossy compression schemes,
allows for the relaxation of certain assumptions, and 
offers performance guarantees. 
We show for Gaussian-distributed initial states with entropy-coded scalar quantization and vector 
quantization that the coding rates for  distributed averaging 
 can be optimized through generalized geometric programming. 
In the absence of side information from past states,
 this approach finds a rate allocation 
over nodes and iterations that minimizes the aggregate 
coding rate  required to achieve a target mean square error within 
a finite run time. 
Our rate allocation is compared to 
some of the prior art through numerical simulations.  
The results motivate the incorporation of side-information through 
differential or predictive coding to improve rate-distortion 
performance.
\end{abstract}


\section{Introduction}
The proliferation of wireless sensors and 
large distributed data sets in recent years has provided significant 
motivation for the development of distributed computing methods. 
In many distributed computing settings, 
it is necessary to compute a function of 
data that may be dispersed among a number of computing nodes. 
Examples include wireless sensor networks
(WSNs), where each agent 
observes a different measurement of a physical process, and
large-scale server farms, where the size of the data set requires distributed 
storage~\cite{NoorshamsWainwright2011}. The class of distributed algorithms 
considered in this paper computes these functions using only interactions
among local subsets of 
the network nodes. One popular approach to distributed function computation, 
which has many variants, is {\em consensus}~\cite{Olfati-Saber2007,Dimakis2010}. 
Consensus has found applications in a wide variety of settings, 
including distributed swarm control, sensor fusion, optimization, filtering, 
environmental monitoring, and distributed 
learning~\cite{Olfati-Saber2007,Nokleby2013,Raja2016}. 

As a motivating example for our lossy compression
framework, we consider the application of distributed averaging in 
representation learning for the internet of things 
(IoT). Distributed average consensus helps different nodes share intermediate 
results within a distributed version of the power 
method~\cite{Scaglione2008},\footnote{ This distributed 
power method helps find the dominant singular vector of a matrix when batches of its 
columns are stored among different nodes of a network.} which can be used 
in distributed dictionary learning schemes such as cloud K-SVD~\cite{Raja2016}. Take face 
recognition as an example. Powered by IoT, a camera network can gather a large 
number of face images and train a face recognition model using dictionary
learning~\cite{Wright2009}. If these cameras can share the data they gathered 
and re-train their face recognition model, then they can improve their 
performance over 
time. In fully connected networks, the cameras would easily share newly gathered 
images and re-train their face recognition model efficiently. However, in IoT, 
the network is often ad-hoc and may have sparse connectivity. Furthermore, in 
 dictionary learning for images, large quantities of data 
must be communicated among nodes, which strains the energy constraints of IoT 
devices~\cite{pottie2000}. 

In this work, we propose a scheme that offers the potential
to balance the competing  metrics of communication load, estimation error, and
execution time for such IoT applications. After running for a finite number of
iterations $T < \infty$, average consensus will produce an estimate of the mean that has
some associated mean square error (MSE), $\MSE{T}>0$. If the internode messages are
quantized, this process will require the transmission of a total 
or aggregate coding rate,
$\Ragg > 0$. Our  goal is to minimize the
communication load, measured by $\Ragg$, subject to a given number of iterations
to be carried out ($T$) and a desired final accuracy, $\MSE{T}$. To perform this
minimization, we use the theory of generalized geometric
programming (GGP)~\cite{BoydVandenberghe2004,BoydKimVandenbergheHassibi2007} to
 formulate a convex program that can find the global optimum solution for
certain lossy compression schemes.

\subsection{Prior art}
Many early works on consensus assumed that the nodes could communicate 
real-valued data to one another~\cite{Frasca2008}. In realistic scenarios, 
the nodes must communicate within bandwidth and energy constraints, 
which can have a significant impact on the convergence of distributed 
averaging algorithms. Although many papers have been published on quantized 
consensus in recent years 
(e.g.,~\Cite{CarliBulloZampieri2009,Thanou2013,YildizScaglione2008,
YildizScaglione2008b,Frasca2008,RajagopalWainwright2011,
NoorshamsWainwright2011,Nokleby2013,HuangHua2011}),\footnote{For a more thorough
literature review, see Pilgrim~\cite{PilgrimThesis2017}.}
a large portion considers trade-offs among run time, communication load, and
final accuracy 
without formulating the problem as  constrained optimization. Early 
publications on quantized consensus (e.g.,~\Cite{XiaoBoydKim2007,
Frasca2008}) show that introducing perturbations of constant variance 
(such as quantization error) into the traditional consensus state update 
prevents convergence due to the limited precision of the quantizer. Due to 
the difficulties associated with quantization error, many 
works~(e.g., \Cite{CarliBulloZampieri2009,Li2011,Rego2015,Thanou2013,
YildizScaglione2008b,YildizScaglione2008}) address the incorporation of dynamic coding
strategies into consensus protocols. However, few prior 
schemes explicitly consider the rate-distortion (RD) trade-off, and they
instead offer 
heuristics to optimize their respective performance metrics.

Several papers~(e.g., \cite{CarliBulloZampieri2009,Li2011,Rego2015}) consider the 
possibility of using differential coding (also called difference 
quantization)~\cite{GershoGray1993} with a shrinking quantization range 
during the transmission of messages  between nodes. Of these, 
Thanou \etal~\cite{Thanou2013} demonstrated lower 
MSE than previous works in this area (e.g., \Cite{Li2011,
CarliBulloZampieri2009}) with equal communication load. Although these 
papers~\cite{Thanou2013,Li2011,CarliBulloZampieri2009} exploit side-information 
in their coding strategies, they study the simple fixed-rate uniform 
quantizer and do not make 
effective 
use of lossy compression to balance the trade-offs among $T$, $\MSE{T}$, and 
$\Ragg$.

Yildiz and Scaglione~\cite{YildizScaglione2008b,YildizScaglione2008}, unlike 
other authors, explicitly considered the RD trade-off to achieve an asymptotic 
MSE value in consensus for the case of Gaussian 
initial states. They proposed schemes based on 
differential~\cite{YildizScaglione2008b}, predictive, and Wyner-Ziv 
coding~\cite{YildizScaglione2008}. Despite their sophisticated coding 
approaches, the state update step they used can only provide 
bounded steady-state error in the limit of many consensus iterations,
 so that 
$\lim_{T\rightarrow\infty} \MSE{T} > 0$~\cite{YildizScaglione2008}.

In addition to these dynamic coding
strategies~\cite{Li2011,CarliBulloZampieri2009,Thanou2013,YildizScaglione2008},
a few works \cite{Nokleby2013,HuangHua2011} consider optimization strategies for
energy consumption in wireless sensor networks. However, Nokleby
\etal~\cite{Nokleby2013} require a specific topological evolution of the
network. Huang and Hua~\cite{HuangHua2011} keep the coding rates for their
fixed-rate uniform quantizers constant across both nodes and iterations, which
does not fully explore the  more general space of node- and
time-varying rates.

A handful of works~(e.g., \Cite{AyasoShahDahleh2010,XuRaginsky2014,SuGamal2010,
YangGroverKar2016}) analyze consensus from the viewpoint of information theory. 
Yang \etal~\cite{YangGroverKar2016} considered RD bounds for data aggregation, 
in which data is routed through a tree network to a fusion center, and 
consensus, in which each node forms an estimate of the desired quantity. 
Although Yang \etal~\cite{YangGroverKar2016} provided bounds on the RD 
relationship for consensus in trees and proved the achievability of the 
derived bounds, their analysis is limited to the setting where the network is 
tree structured. Often it is beneficial to consider more flexible topologies, 
such as random geometric graphs, which have been used to model 
WSNs~\cite{BoydMixingTimes}. In general, random geometric graphs and their 
real-world WSN counterparts have loops.

\subsection{Our contributions}
This paper presents a framework for attaining an estimate of the network sample
mean at each node, within a desired average level of accuracy, with finite run
time and minimal total communication cost (measured by $\Ragg$) using
either deterministic or dithered quantization. Our 
framework is informed by the results of RD theory~\cite{Cover06} and
convex optimization~\cite{BoydVandenberghe2004}. In the plethora of literature
we surveyed, 
{\em a crucial problem that has not been addressed is the optimization
of quantization schemes for finite run-time without a certain set of limiting
assumptions}.  
Prior works 
restrict the topology to trees~\cite{YangGroverKar2016},
 assume a certain form for the rate/distortion
sequences~\cite{YildizScaglione2008,YildizScaglione2008b}, 
restrict the rates to be the same at each node and/or
iteration~\cite{HuangHua2011,Thanou2013,YildizScaglione2008b}, or 
use bounds on the MSE or asymptotic MSE values, rather than exact MSE
quantities~\cite{HuangHua2011,YildizScaglione2008b}, in their analyses. The key
contribution of this work is the use of 
GGP~\cite{BoydKimVandenbergheHassibi2007} to minimize the
communication load subject to an accuracy constraint, which avoids the
previously discussed restrictions. The advantages of our approach are ({\em i})
ignorance about the parametric form of the allocated rates, which avoids the
introduction of unjustified or unnecessary assumptions;
({\em ii}) support for
different rates at each node and iteration of 
distributed average consensus; 
and ({\em iii}) optimization with
respect to  exact MSE constraints (rather
than bounds) for finite iteration count.

\subsection{Notation}
We denote the positive-valued subset of a set $\mathcal{S}$ by $\mathcal{S}_{>0}$ and
the nonnegative-valued subset by $\mathcal{S}_{\geq 0}$. The integers are denoted by
$\mathbb{Z}$, and the real numbers by $\reals$. Vectors are written in boldface
lowercase letters (e.g., $\mathbf{x}$). Matrices are written in boldface capital
letters (e.g., $\mathbf{A}$). The $(i,j)\th$ element of matrix $\A$ is written
$\left[\A\right]_{ij}$. Quantities that vary with time are written as
functions of time (e.g., $x(t)$). The $\ell_p$-norm is written
$\norm{\cdot}_p$, matrix transpose is written $\{\cdot\}\tp$, and matrix inverse
is written $\{\cdot\}^{-1}$. The expectation operator is written as $\E{\cdot}$, the mean of a time-dependent 
random variable $\x(t)$ is written $\mean{\x}{t}$, and its covariance is denoted by
$\Sig{\x}{t}$. The Kronecker product of two matrices $\A, \mathbf{B}$ is 
written $\A \otimes \mathbf{B}$. A diagonal matrix is compactly specified as
\begin{align}
    \diag 
    \begin{bmatrix} 
        x_1 \\
        \vdots \\
        x_n 
    \end{bmatrix} = 
    \begin{bmatrix}
        x_1 & \cdots & 0 \\
        \vdots & \ddots & \vdots \\
        0 & \cdots & x_n
    \end{bmatrix}.
\end{align}

\section{Problem formulation}
\subsection{System model}

In this paper, communication links are bidirectional, and we
model the network as an undirected graph, $\mathcal{G} = \{\mathcal{V}, \mathcal{E}\}$, 
comprised of a set of $m$ vertices (nodes) 
$\mathcal{V}$ and a set of edges $\mathcal{E}$ between pairs of 
vertices~\cite{GraphsOptimizationAndAlgos}. Because the communication links are 
bidirectional, each edge $(i,j) \in \mathcal{E}$ is represented as an 
unordered pair of vertices, $i, j\in\mathcal{V}$~\cite{XiaoBoyd2004}.

In the simplest case of the consensus problem, each node $i \in \{1,\ldots,m\}$
has an initial scalar quantity $z_i(0) \in \reals$, and the goal is to have all
nodes of the network agree upon the sample mean of these quantities by
iteratively exchanging messages with their neighbors~\cite{XiaoBoyd2004}. The
quantities $z_i(t)$, $t \geq 0$, will be referred to as ``states,''
which in this paper are assumed to be real-valued scalar random variables (RVs)
with  joint Gaussian distribution for $t=0$. More
formally, let the (discrete) iteration index be a nonnegative integer, $t \in 
\mathbb{Z}_{\geq 0}$. At $t=0$, the states $\{z_i(t)\}_{i=1}^m$ are the initial 
values to be averaged by the consensus algorithm. For $t \geq 1$, the state 
$z_i(t)$ represents the estimate of the sample average $\widebar{z} := 
\frac{1}{m} \sum_{i=1}^m z_i(0)$ at node $i$. The objective of consensus is for 
the state $z_i(t)$ to eventually equal the sample mean of the initial states: 
$\lim_{t\rightarrow\infty} z_i(t) = \widebar{z}$, $\forall i \in 
\{1,\ldots,m\}$~\cite{XiaoBoyd2004}. In this paper,
we restrict our attention to deterministic, synchronous-update consensus
algorithms. We assume the following: 
({\em i}) the communication link topology
of the network is fixed and does not change with time; 
({\em ii}) at each iteration, every node 
exchanges messages with only its neighbors;
({\em iii}) communication channels are noiseless; 
({\em iv}) initial states of all nodes are 
Gaussian with a known joint distribution;
({\em v}) each node can use a different rate at each 
consensus iteration;
({\em vi}) internode messages are broadcast to all neighbors at once; and
({\em vii}) states are stored with infinite precision, but communicated with 
limited precision. The last point is well-motivated for nodes 
with 32- or 
64-bit floating-point support. 

Given the above assumptions on communication, one  popular 
algorithm
for consensus relies on linear updates~\cite{XiaoBoyd2004,Olfati-Saber2007}.
Each node updates 
its state by forming a weighted sum of its own state with those of its 
neighbors~\cite{XiaoBoyd2004}, 
\begin{align}
\label{eq:node-update}
z_i(t+1) &= w_{ii} z_i(t) +  \sum_{j \in \mathcal{N}_i} w_{ij} z_j(t) ,
\end{align}
where $w_{ij} > 0$, $\forall i,j$, $\sum_{k=1}^m w_{ik}$ $ = \sum_{k=1}^{m} w_{kj}$ $= 1$, and
$\mathcal{N}_i = \{j | (i,j) \in \mathcal{E} \}$ denotes the 
neighborhood of node $i$. The weights $w_{ij}$ are designed such that 
$\lim_{t\rightarrow \infty} z_i(t) = \widebar{z}$~\cite{XiaoBoyd2004}.
The above update equation~\eqref{eq:node-update} can be written 
as~\cite{XiaoBoyd2004}
\begin{align}
\label{eq:update-mv}
\z{t+1} = \W{} \z{t} .
\end{align}
The asymptotic convergence condition is then
$\lim_{t \rightarrow \infty} \z{t} = \frac{1}{m} \ones \ones \tp \z{0} = 
\widebar{z}\,\ones$. The interested reader is referred to Xiao and 
Boyd~\cite{XiaoBoyd2004} for a study of weight matrix design. 

When quantization error is
present within the internode messages, the simple linear
iteration~\eqref{eq:update-mv} is not guaranteed to
converge~\cite{XiaoBoydKim2007}. Instead, we use the modified
iteration used by Frasca \etal~\cite{Frasca2008}, which allows the sample
average to be preserved in the presence of quantization errors. 


Let $Q: \reals^m \to \mathcal{X}^m$ represent quantization to a finite set of
representation levels $\mathcal{X}^m \subset \reals^m$ (i.e., $\q{\z{t}} =
[Q_1(\zmarg{1}{t}), \ldots, Q_m(\zmarg{m}{t})]\tp$). The associated
quantization error is given by $\qes{t} := \q{\z{t}}-\z{t}$. We define the
distortion at node $i$ and iteration $t$ as
$D_i(t) := \E{\epsilon_i^2(t)}$. The subscripts on $Q$
indicate that each node can use a different quantizer in general. 
To allow the algorithm to converge to
zero steady-state estimation error, we use the update proposed by 
Frasca \etal~\cite{Frasca2008}, which is
\begin{align}
\label{eq:lossy-iteration}
\z{t+1} &= \z{t} + \left(\W{} - \I\right)\q{\z{t}} ,
\end{align} 
where $\I$ is the identity matrix.
The key advantage of this update is that the average $\frac{1}{m} \sum_{i=1}^m
z_i(t)$ of the states $z_i(t)$ is preserved at each step $t$, despite
quantization error~\cite{Frasca2008}. Because the average is preserved at each iteration, the estimation error from 
the average consensus state $\widebar{z}\ones$ is~\cite{Frasca2008}
\begin{align}
\label{eq:est-error}
\diff{t} = \M \z{t}.
\end{align}

The MSE at node $i$ corresponding to an estimation error $\diff{t}$ at
iteration $t$ is given by 
\begin{align}
    \label{eq:initial-node-mse-def}
\text{MSE}_i(\Dv, t) := \E{e_i^2(t)},
\end{align}
where $\Dv$ is a vector of all distortions introduced by all nodes throughout
the consensus process.
The average MSE across the network at the end of iteration $t$ is given by
\begin{align}
    \label{eq:initial-mse-def}
    \text{MSE}(t) := \frac{1}{m} \sum_{i=1}^m \text{MSE}_i(t).
\end{align}

The communication cost of consensus becomes substantial when nodes
exchange vector states $\mathbf{z}_i(t) \in \reals^n$, rather than scalars.
In this case, all entries can be collected into a vector
$\ze{t} := \left[\mathbf{z}\tp_1(t), \cdots, \mathbf{z}\tp_m(t) \right]\tp \in
\reals^{mn},$
and a matrix is defined $\boldsymbol{\Omega} := \W{} \otimes \I_n \in
\reals^{mn \times mn}$~\cite{HuangHua2011}, 
so that \eqref{eq:lossy-iteration} becomes
\begin{align}
    \label{eq:lossy-vector-iteration}
    \ze{t+1} = \ze{t} + \left( \Om - \I_{mn} \right) Q\left(\ze{t} \right).
\end{align}

This work assumes that $\ze{t}$ has a known joint Gaussian
distribution, and that the $\mathbf{z}_i(t)$ are distributed such that
$\boldsymbol{\mu}_{\mathbf{z}_i(t)} = \mu_{\mathbf{z}_i(t)} \ones$, and 
the diagonal of $\Sig{\mathbf{z}_i}{t}$ is $\sigma_i^2(t) \ones$. 
This means that the 
marginal mean and variance can be extracted from any of the
elements corresponding to $\mathbf{z}_i(t)$ from the joint mean
$\mean{\boldsymbol{\zeta}}{t}$ and joint covariance
$\Sig{\boldsymbol{\zeta}}{t}$, respectively. The marginal means and variances
can thus be derived in our case from~\eqref{eq:lossy-iteration}, without 
considering the higher-dimensional~\eqref{eq:lossy-vector-iteration}. The
following statistical analysis will focus on this scalar case.

\subsection{Main objective}
At every iteration $t \in \{0,\ldots,T-1\}$ (where $T$ is the total number of iterations) of the consensus process, each node
$i \in \{1,\ldots,m\}$ uses a rate $R_i(t)$ to encode its state for
transmission to neighboring nodes. This 
rate is the average number of bits used per symbol. That is, if node
$i$ sends a length-$l_i(t)$ binary encoding of its scalar state $z_i(t)$ to node $j$, then
the corresponding rate is given by $R_i(t)=\E{l_i(t)}$, where the expectation
is taken over the distribution of $z_i(t)$. In general, $R_i(t)$ can vary
across both nodes and iterations, so that it is not necessary that
$R_i(t)=R_j(s)$ for any $i\neq j, t\neq s$.  We simplify notation by defining 
the {\em rate vector}
\begin{align}
\label{eq:r-vector}
\Rhat \!:=\! \left[R_1(0),...,\!R_m(0),...,\!R_1(T\!-\!1),..., \!R_m(T\!-\!1) \right]\tp\!\!\!\!.
\end{align}
Denote the distortions per entry, $D_i(t):=\E{\epsilon_i^2(t)}$,
incurred using rates $\Rhat$ by the {\em distortion
vector}
\begin{align}
\label{eq:d-vector}
\Dv \!:=\! \left[D_1(0),...,\!D_m(0),...,\!D_1(T\!-\!1),...,\!D_m(T\!-\!1) 
\right]\tp\!\!\!\!.
\end{align}
One key quantity we use to determine the cost of running the consensus process 
is the {\em aggregate coding 
rate}~\cite{ZhuBaronMPAMP2016ArXiv,ZhuPilgrimBaron2017}:
\begin{align}
\label{eq:ragg-def}
R_{\text{agg}} := \sum_{t=0}^{T-1} \sum_{i=1}^m R_i(t),
\end{align}
which represents the total rate used over the $T$ iterations of the 
consensus algorithm by all $m$ nodes of the network.

Our main objective is to derive minimization strategies for 
\begin{align}
\label{eq:comm-cost}
C(\Rhat,T):= R_{\text{agg}}
\end{align}
for fixed- and variable-length codes for Gaussian-distributed sources using a
variety of quantizers. 

To efficiently encode the data stored across the network,
it is necessary to know the distribution of $z_i(t)$ for all $i \in
\{1,\ldots,m\}$ and $t\geq 0$.
To determine these distributions from~\eqref{eq:lossy-iteration} and the
distributions of the initial states
$\{z_i(0)\}_{i=1}^m$ is difficult in general. Instead, we propose
an optimization scheme for entropy-coded uniform scalar quantization
(ECSQ)~\cite{GershoGray1993} of stationary Gaussian states and RD-optimal vector
quantization (VQ)~\cite{Cover06} of memoryless Gaussian-distributed 
states.\footnote{For the scalar quantization schemes, we assume that the state 
of each node is a sample from a stationary, ergodic Gaussian random process. 
We expect the performance of ECSQ to be the same as in the memoryless case.} 
Because~\eqref{eq:lossy-iteration} consists of a linear combination of jointly
    Gaussian RVs and independent quantization errors, it can be proven that the
states will remain Gaussian for all $t\geq 0$~\cite{PilgrimThesis2017}, and thus
the mean and covariance of $\z{t}$ are sufficient to describe its distribution.

It can be shown that, for additive quantization noise and symmetric weight
matrices~\cite{PilgrimThesis2017}:
\begin{align}
\label{eq:se-data-mean-simple}
\mean{\mathbf{z}}{t+1} = \W{} \mean{\mathbf{z}}{t},
\end{align}
\begin{align}
\label{eq:se-error-mean-simple}
\mean{\mathbf{e}}{t+1} = \M \W{} \mean{\mathbf{e}}{t}, 
\end{align}
\begin{align}
\label{eq:se-data-simple}
\Sig{\mathbf{z}}{t+1} &= \W{} \Sig{\mathbf{z}}{t} \W{} + \WsubI \Sig{\qesu}{t} \WsubI,
\end{align}
\begin{align}
\label{eq:se-error-simple}
\Sig{\mathbf{e}}{t} = \M \Sig{\mathbf{z}}{t} \M ,
\end{align}
where $\mathbf{e}(t)$ is the estimation error~\eqref{eq:est-error} and the
covariance
$\Sig{\qesu}{t}$ is diagonal, $\Sig{\qesu}{t}=
\diag [D_1(t),\cdots,D_m(t)]\tp$. 
Using these definitions, we present the following mathematical relationships, 
which we term the {\em state evolution equations}. These equations allow us to 
perform the optimization of the rate vector $\Rhat$ using the cost 
function~\eqref{eq:comm-cost}.
The marginal source variance $\margvar{i}{\Dv,t}$ at node $i$ and 
iteration $t$ is given by
\begin{align}
\label{eq:se-var}
\margvar{i}{\Dv, t} &= \left[ \Sig{\mathbf{z}}{t} \right]_{ii},
\end{align}
the MSE at node $i$ and iteration $t$ is given by
\begin{align}
\label{eq:se-mse}
\mathrm{MSE}_i(\Dv, t) &= \left[  \Sig{\mathbf{e}}{t} + 
\mean{\mathbf{e}}{t}\meant{\mathbf{e}}{t} \right]_{ii} ,
\end{align}
and the average MSE across the network at iteration $t$  is
\begin{align}
\label{eq:se-mse-net}
\MSE{\Dv, t} &= \frac{1}{m} \tr{\Sig{\mathbf{e}}{t} + 
\mean{\mathbf{e}}{t}\meant{\mathbf{e}}{t} },
\end{align}
where $\tr{\cdot}$ denotes the trace of a matrix.

\subsection{Rate-distortion theory}
For encoders operating on real-valued sources, the quantization process 
necessarily introduces a certain expected distortion $D$ into their 
representation of the input signal~\cite{GershoGray1993}. This distortion can be 
quantified using a number of metrics, but for the purpose of this paper, we use 
the square error
\begin{align}
    \delta(\z{t},\hat{\mathbf{z}}(t)) = \norm{\z{t} - \hat{\mathbf{z}}(t)}_2^2,
\end{align} 
so that the expected distortion per
node per dimension is given by $D =
\frac{1}{m} \E{\norm{\z{t}-\hat{\mathbf{z}}(t)}_2^2}$~\cite{GershoGray1993}. In 
general, using a higher coding rate $R$ results in a lower distortion $D$, with 
the drawback of greater communication load. RD theory~\cite{Cover06} quantifies 
the best possible trade-off between coding rate and distortion. The minimum 
coding rate $R$ required for any compression scheme to produce an expected 
distortion less than or equal to a particular value $D$ is given by the RD 
function $R(D)$~\cite{Cover06}.

\section{Rate allocation via GGP}
The key insight of this work is the ability to pose the optimization of the
rate vector $\Rhat$ as a GGP, for which
the global optimum can be found~\cite{BoydKimVandenbergheHassibi2007}.   
The resulting scheme finds an efficient 
rate vector $\Rhat$ that achieves a target value of
$\MSE{\Dv,T}$, given by~\eqref{eq:se-mse-net}.\footnote{Note that the final network MSE 
corresponds to iteration index $T$, and not $T-1$. This is because $\MSE{\Dv,t}$ 
is the network MSE {\em after the end of} $t$ iterations, or {\em before the 
execution of the $(t+1)\th$ iteration}.} 

When a particular quantizer is used in our problem, it will often have an RD 
performance trade-off that differs from $R(D)$, which is a bound on the best possible 
performance~\cite{GershoGray1993}. In this paper, we term such a trade-off curve 
for a particular practical quantizer an {\em operational RD relationship}.

For ECSQ and uniform quantization
followed by fixed-rate coding in the case of  Gaussian sources, the 
operational RD relationship in the high-rate regime is
\begin{align}
\label{eq:pattern}
R(D) \approx \begin{cases}
\dfrac{1}{2} \log_2 \left(\dfrac{\sigma^2}{D} \right) + R_c, \fourspace &D 
\in (0, \sigma^2 D_{\text{max}}] \\
0, \fourspace &\text{otherwise}
\end{cases},
\end{align}
where $\sigma^2$ represents the variance of the data to be encoded, and
$R_c$ and $D_{\text{max}}$ are constants~\cite{GishPierce68}. In
some cases, such as infinite-dimensional VQ with memoryless Gaussian sources and
dithered~\cite{Lipshitz1992} scalar uniform quantization~\cite{ZamirFeder96},
the relationship~\eqref{eq:pattern} holds for all rates. 

The source variance $\margvar{i}{\Dv, t}$ is a function of the
initial state vector covariance $\Sig{\mathbf{z}}{0}$ (see~\eqref{eq:se-data-simple})
and the distortion vector $\Dv$ in~\eqref{eq:d-vector}; it evolves as
described by~\eqref{eq:se-data-simple}. The operational RD relationship at all 
nodes $i \in \{1,\ldots,m\}$ and iterations $t\in\{0,\ldots,T-1\}$ can be 
expressed as
\begin{align}
\label{eq:rd}
R_i(\Dv, t) &=\frac{1}{2}  \log_2 \left( \max \left\{ 
\frac{\margvar{i}{\Dv,t}}{\Dmarg{i}{t}}, 2^{-2R_c} \right\} \right) + R_c . 
\end{align}
The $\max$ in~\eqref{eq:rd} encapsulates the saturation of the RD
relationship at $R=0$ in~\eqref{eq:pattern}. Given $T$ iterations,
the goal is to minimize the aggregate coding rate~\eqref{eq:ragg-def}, subject
to a constraint on the final MSE, $\MSE{\Dv,T} \leq
\text{MSE}^*$~\eqref{eq:se-mse-net}. For a target MSE of 
${\rm{MSE}}^*$, the minimum number of iterations required to achieve
that MSE, $T_{\text{min}}= \argmin_T \left \lbrace T \,\, \vert \,\,
\MSE{\Dv,T} < {\rm{MSE}}^*, \Dv = \mathbf{0} \right \rbrace$, can be readily obtained using the 
state-evolution equations~\cref{eq:se-data-mean-simple,eq:se-error-mean-simple,eq:se-data-simple,eq:se-error-simple}. 
More formally, using the operational RD relationship~\eqref{eq:rd}, the 
optimization problem is
\begin{align}
\underset{\Dv}{\text{minimize}} \sum_{t=0}^{T-1} \sum_{i=1}^m \frac{1}{2} \log_2\left( \max \left\{ \frac{\margvar{i}{\Dv,t}}{\Dmarg{i}{t}}, 2^{-2R_c} \right\} \right) + R_c,
\end{align}
subject to the constraints
\begin{align}
\MSE{\Dv, T} \leq \text{MSE}^*, \label{eq:constone}
\end{align}
\begin{align}
\Dmarg{i}{t} > 0, \eightspace \forall i,t \label{eq:consttwo}.
\end{align}
Note that the above optimization is equivalent to
\begin{equation}
\begin{aligned}
\label{eq:cfun}
&\underset{\Dv}{\text{minimize}}
& & \ln \left( \prod_{t=0}^{T-1} \prod_{i=1}^m \max 
\left\{ \frac{\margvar{i}{\Dv,t} }{\Dmarg{i}{t}}, 2^{-2R_c} \right\} 
\right), \\ \vspace{12pt}
& \text{subject to}
&& \hspace{54pt}\text{\eqref{eq:constone}~and~\eqref{eq:consttwo}}. 
\end{aligned}
\end{equation}
We will now introduce the concept of GGP and show that the 
optimization~\eqref{eq:cfun} reduces to such a problem.

\subsection{Basics of GGP} The following 
information can be found in Boyd and Vandenberghe~\cite{BoydVandenberghe2004}. 
For this subsection, we stay close to the authors' original notation. In the 
language of geometric programming, a function of the form
\begin{align}
f(x) = c x_1^{a_1}x_2^{a_2}\cdots x_n^{a_n}, \fourspace c >0, 
\fourspace x_i > 0, \fourspace a_i \in \reals, \fourspace \forall i,
\end{align}
is called a {\em monomial}~\Cite{BoydVandenberghe2004,
BoydKimVandenbergheHassibi2007}. Similarly, a function of the form
\begin{align}
f(x) = \sum_{i=1}^k g_i(x_1,\ldots,x_n),
\end{align}
is a {\em posynomial}~\Cite{BoydVandenberghe2004,
BoydKimVandenbergheHassibi2007}, where $g_i(x_1,\ldots,x_n)$ are monomials. 
Generalized posynomials are functions formed from posynomials by operations 
including addition, multiplication, and 
maximum~\cite{BoydKimVandenbergheHassibi2007}.

\begin{figure*}
\vspace*{3mm}
\floatbox[{\capbeside\thisfloatsetup{capbesideposition={right,top},
capbesidewidth=44mm}}]{figure}[\FBwidth]
{\caption{\small
Optimal rates and MSE sequences from the solution 
        of~\eqref{eq:cfun-final} ($T=5$, $\rho_c=0.35$, $\sigma_x^2=1$, 
    $\sigma_n^2=0.5$, $m=20$). {\em Left:} Optimal rate sequences for~\eqref{eq:cfun-final}. {\em Right:} 
Optimal rate sequences for~\eqref{eq:cfun-mod}. The rates are plotted against iteration indices, 
and each line represents the rates used by a different sensor. Note that because 
variable-length coding is used~\cite{GershoGray1993}, the rates can be 
non-integer-valued.}\label{fig:varD-optimal-rates}}
{\includegraphics[width=61mm]{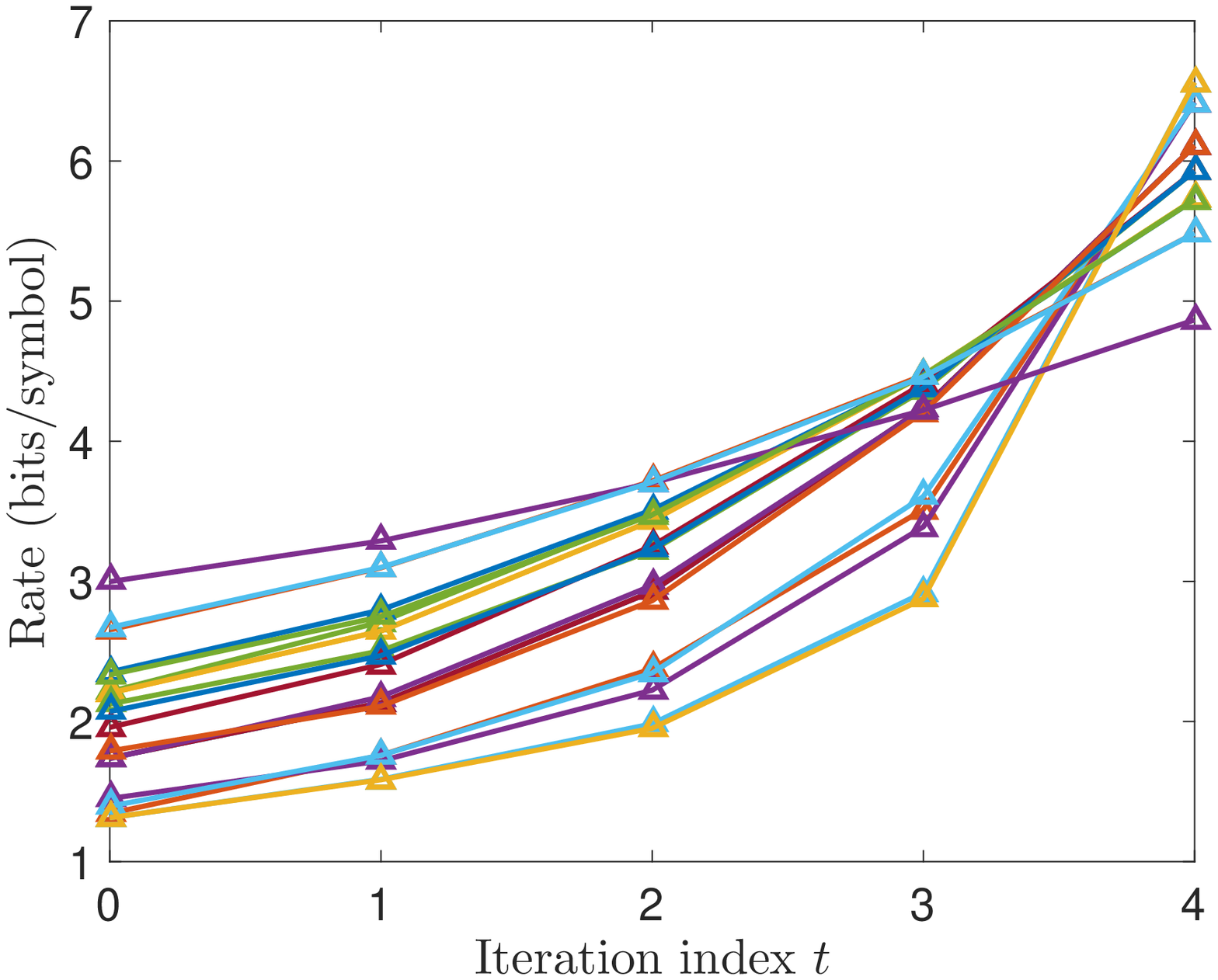}
	\hspace{2mm}
{\includegraphics[width=61mm]{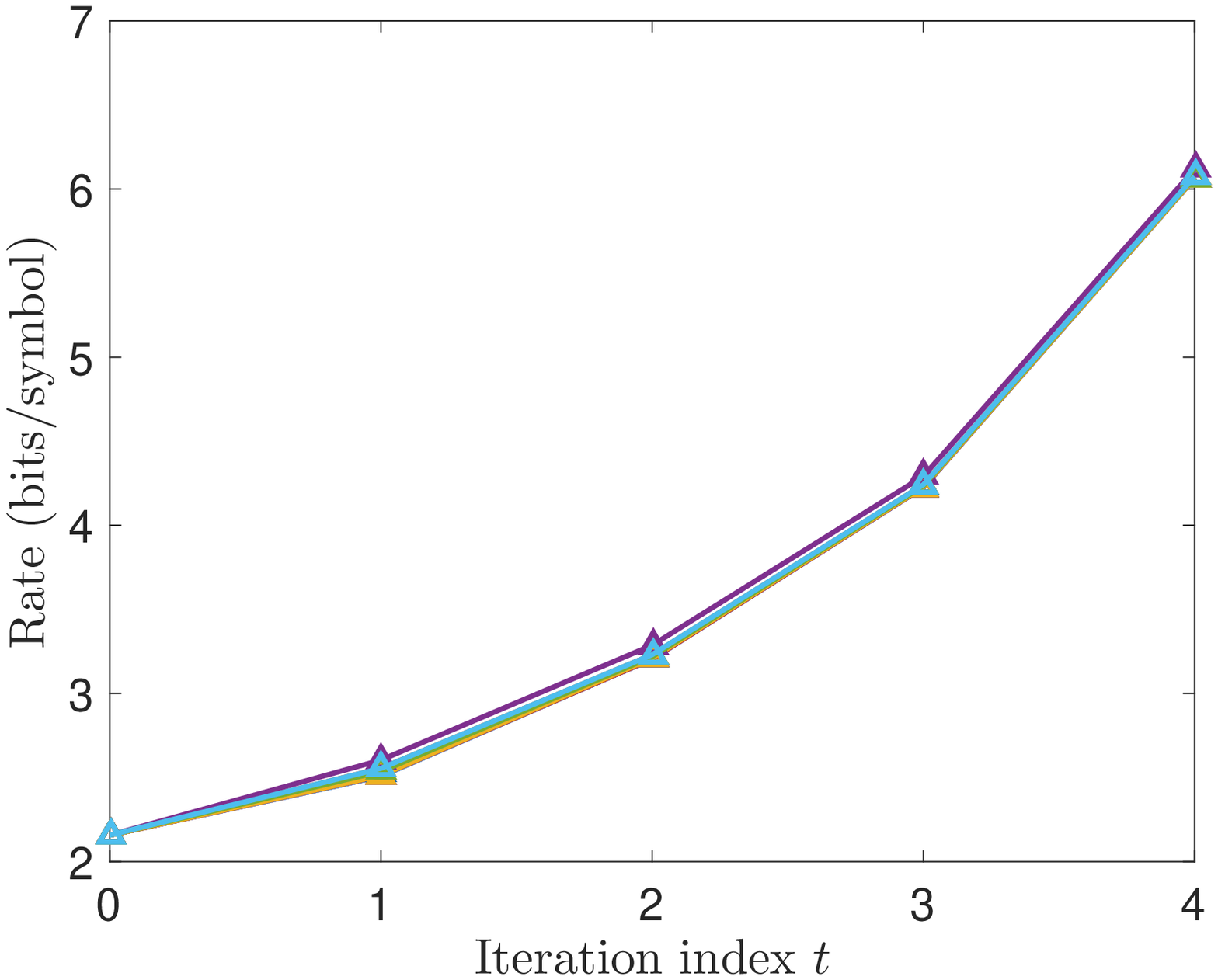}}}
\vspace{-6mm}
\end{figure*}

A standard inequality-constrained GGP has the form
\begin{equation}
\begin{aligned}
&\underset{x_1,\ldots,x_n}{\text{minimize}}
& & C(x_1,\ldots,x_n), \\
& \text{subject to} 
& & \fourspace f_i(x_1,\ldots,x_n) \leq 1, \eightspace \forall i \in \{ 1,\ldots,n_f \}, \\
& & & \fourspace g_i(x_1, \ldots, x_n) = 1, \eightspace \forall i \in \{ 1,\ldots,n_g \}, \\
& & & \fourspace x_i > 0 \fourspace \forall i \in \{1, \ldots, n \},
\end{aligned}
\end{equation}
where the cost $C(x_1, \ldots, x_n)$ and all the inequality constraints 
$f_i(x_1, \ldots, x_n)$ are generalized posynomials, and all the equality 
constraints $g_i(x_1, \ldots, x_n)$ are 
monomials~\cite{BoydKimVandenbergheHassibi2007}. 
Applying some function transformations, 
GGPs can be cast in convex form and 
efficiently solved 
numerically~\cite{BoydKimVandenbergheHassibi2007}.

\subsection{Generalized posynomial form of cost function}

Applying a monotone increasing function to the cost function~\eqref{eq:cfun}
results in an equivalent problem 
to~\eqref{eq:cfun}~\cite{BoydVandenberghe2004}, so we apply the 
exponential function to obtain
\begin{equation}
\begin{aligned}
\label{eq:cfun-final}
&\underset{\Dv}{\text{minimize}}
& & \prod_{t=0}^{T-1} \prod_{i=1}^m \max \left\{ \frac{\margvar{i}{\Dv,t} }{\Dmarg{i}{t}}, 2^{-2R_c} \right\}, \\
\vspace{12pt}
& \text{subject to} 
& & \hspace{48pt}\text{\eqref{eq:constone}~and~\eqref{eq:consttwo}}. 
\end{aligned}
\end{equation}

The above optimization problem~\eqref{eq:cfun-final} is a
GGP, which is formally shown in Pilgrim~\cite{PilgrimThesis2017}.

Two aspects of the above optimization~\eqref{eq:cfun-final} should be
highlighted. Because the constraints are allowed to be generalized posynomials,
one could also optimize with respect to a constraint on the maximum node MSE,
for example,
\begin{align}
\max_i \,\, \{\text{MSE}_i(\Dv,T)\}_{i=1}^m \leq \text{MSE}^*,
\end{align} 
or constraints on each of the node MSE values, 
\begin{align}
\text{MSE}_i(\Dv,T) \leq \text{MSE}^*_i, \fourspace \forall i \in \{1,\ldots,m\},
\end{align}
where $\text{MSE}_i(\Dv,t)$ is defined in~\eqref{eq:se-mse}. Also, in its most 
general form, the optimization allows each node to use a different rate or 
distortion. In the interest of designing a distributed protocol (or for 
computational efficiency), one may wish to constrain the rates or distortions at 
each node to be the same. The constraint that all distortions be the same is a 
straightforward modification of~\eqref{eq:cfun-final} and is also a GGP.

\subsection{Constant distortion simplification}
Solving the exact optimization problem~\eqref{eq:cfun-final} naively requires
explicit representation of all $mT$ distortions $D_i(t)$ and all the
coefficients of the log-sum-exp (LSE) model\footnote{GGPs are converted to
convex LSE form for solution~\cite{cvx,gb08}.} required to compute
$\text{MSE}_i(\Dv,t)$ and $\margvar{i}{\Dv,t}$ from $\Dv$. The result of this
explicit representation is large computational time and memory
complexity. In this section, we explore a simplification
of~\eqref{eq:cfun-final}  to combat these issues. 

In practice, as the network grows (specifically, $m > 20$
and $T \geq 6$), the memory and time requirements of the
optimization~\eqref{eq:cfun-final} seem to grow quickly. If explicit 
representation of LSE parameters can be avoided, it  is possible to apply
other convex optimization methods without these scaling issues. To provide a
program that is more easily solvable in practice, we 
constrain the distortions to be equal at each node, which is equivalent to 
redefining $\Dv := \left[ D(0),\ldots,D(T-1)\right]\tp$. The optimization is 
then
\begin{equation}
\begin{aligned}
\label{eq:cfun-mod}
&\underset{\Dv}{\text{minimize}}
& & \prod_{t=0}^{T-1} \prod_{i=1}^m  \max{  \left\{ \frac{ \margvar{i}{\Dv, t} }{D(t)}, 2^{-2R_c} \right\}  }, \\
\vspace{12pt}
& \text{subject to} 
& & \hspace{48pt} \text{\eqref{eq:constone}~and~\eqref{eq:consttwo}}.
\end{aligned}
\end{equation}
In the following section, the results of solving the 
simplified problem~\eqref{eq:cfun-mod} are compared to the solutions of the 
exact program~\eqref{eq:cfun-final} and the prior 
art~\cite{Thanou2013,YildizScaglione2008}. Surprisingly, the above simplified 
optimization provides competitive results for random geometric 
networks~\cite{Penrose2003}, with significant reduction in memory and run-time 
requirements. We conclude by noting that in practical implementation, it is anticipated that the
    optimization~\eqref{eq:cfun-final} will be run offline with {\em a priori} 
    knowledge of the network topology and initial state statistics.


\section{Numerical experiments}
In this section, we present numerical results that provide insight into the
optimal rate sequences that result from solving~\eqref{eq:cfun-final}. 
We further compare the performance of the proposed GGP 
optimizations~\eqref{eq:cfun-final} and~\eqref{eq:cfun-mod} to the prior 
art~\cite{YildizScaglione2008,Thanou2013}. To test the effectiveness of the 
proposed approach, we used  the CVX 
toolbox~\cite{cvx,gb08} to solve~\eqref{eq:cfun-final} and~\eqref{eq:cfun-mod}. 

\begin{figure*}
\vspace*{3mm}
\floatbox[{\capbeside\thisfloatsetup{capbesideposition={right,top},
capbesidewidth=44mm}}]{figure}[\FBwidth]
{\caption{\small
{\em Left:} RD trade-off curves for the proposed GGP-optimized 
        ECSQ versus ProgQ~\cite{Thanou2013} and order-one predictive 
    coding~\cite{YildizScaglione2008} ($\rho_c=0.45$, $T=7$). {\em Right:} 
Comparison of the proposed GGP-optimization versus order-one predictive for a 
single realization of the RGG ($\rho_c=0.35$, $T=7$), where both schemes use 
RD-optimal VQ. The corresponding quantization error was simulated by adding 
white Gaussian noise.}\label{fig:tradeoff-curve-ysp1}}
{\includegraphics[width=61mm]{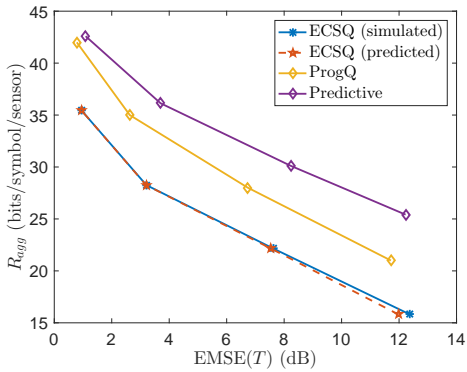}
	\hspace{2mm}
{\includegraphics[width=61mm]{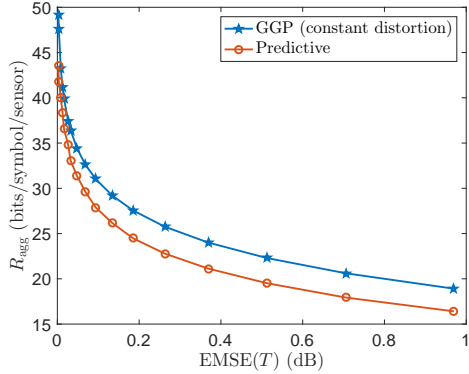}}}
\vspace{-6mm}
\end{figure*}

Due to scaling issues, only the fixed-distortion problem~\eqref{eq:cfun-mod} was 
solved for prior art comparison. The bin size of all fixed-rate uniform 
quantizers was set to 12 times the standard deviation of the data to prevent 
clipping.

In the results presented, the networks were generated by random geometric graph (RGG)
models~\cite{Penrose2003}, and each node state was initialized with the same
independent and identically distributed (i.i.d.) variance-$\sigma_x^2$ zero-mean
Gaussian vector $\x \in \reals^L$ corrupted by a different
variance-$\sigma_n^2$, zero-mean Gaussian noise $\mathbf{n}_i \in \reals^L$, $i
\in \{1,\ldots, m\}$. Incorporating noise is important so that the
nodes have different estimates of the signal $\x$ to average together. The
consensus process averages the states elementwise, so this is the same as
running $L$ trials of consensus on scalar states
$z_i(0)$ at once. The random geometric graphs (RGGs)~\cite{Penrose2003} were
generated on the unit torus (i.e., edge effects were neglected by ``wrapping''
edges of $[0,1]^2$)~\cite{SquareTorus}. The RGGs provide a model of
networks where location determines topology, such as
WSNs~\cite{gupta2000capacity}.
An RGG is one for which each node $V_i \in \mathcal{V}$ is associated with a 
coordinate $\mathbf{v}_i$. For a given connectivity radius $\rho_c$, two nodes 
$V_i, V_j$ are connected if $\norm{\mathbf{v}_i-\mathbf{v}_j}_2 
\leq \rho_c$~\cite{BoydMixingTimes}.

To better understand the structure of the solutions to the optimization
problems~\eqref{eq:cfun-final}~and~\eqref{eq:cfun-mod}, we present some
 simulation results. Each of these results is taken from single
instantiations of the optimization problem (i.e., they are not averaged
over multiple trials).

The optimal rate sequences, $\{R_i(t)\}_{t=0}^{T-1}$, $i \in \{1,\ldots,m\}$, for 
both the variable-distortion \eqref{eq:cfun-final} and 
constant-distortion \eqref{eq:cfun-mod} problems typically exhibit monotonically 
nondecreasing structure, with an increasing rate of change toward the final 
iterations. In the constant-distortion case, the rates $R_i(t) \approx 
\frac{1}{2} \log_2\left( \max \left\{ \frac{\margvar{i}{\Dv, t}}{D_i(t)},
2^{-2R_c} \right\} \right) + R_c$ are 
similar because the ratios $\frac{\margvar{i}{\Dv,t}}{D_i(t)}$ in the Gaussian 
operational RD relationship~\eqref{eq:pattern} are similar across the network. 
Examples of optimal rate sequences are provided for both variants of the 
optimization problem~\cref{eq:cfun-final,eq:cfun-mod} in 
Fig.~\ref{fig:varD-optimal-rates}.

The pattern of these rate sequences is intuitive, and it mirrors the results of Zhu 
and coauthors' study of multiprocessor approximate message 
passing~\Cite{ZhuBaronMPAMP2016ArXiv,ZhuBeiramiBaron2016ISIT,
ZhuPilgrimBaron2017}. As the estimate of the sample mean at each node increases 
in precision, higher-resolution messages must be exchanged among nodes to 
achieve increasing estimation quality. In the case of coding without side 
information, this improving precision requires using larger coding rates in the 
later iterations.

To compare our work to the prior art~\cite{YildizScaglione2008,Thanou2013}, we 
generated 32 RGGs~\cite{Penrose2003} with connectivity radius 
$\rho_c \in \{0.35, 0.45\}$ on a two-dimensional unit torus. For each of these 
networks, consensus was run on 1,000 realizations of the initial states, which 
were length-10,000 i.i.d. Gaussian vectors $\mathbf{z}_i(0) = \x + 
\mathbf{n}_i,$ $\forall i \in \{1,\ldots,m\}$, $\x \sim 
\mathcal{N}(\mathbf{0}, \I), \mathbf{n}_i \sim \mathcal{N}(\mathbf{0}, 0.5 \I)$.
This corresponds to $\text{SNR} := \frac{\sigma_x^2}{\sigma_n^2} = 2$, 
which is 3.01 dB. We simulated ProgQ~\cite{Thanou2013} and order-one 
predictive coding~\cite{YildizScaglione2008}, using initial rates $R_i(0) 
\in \{4,\ldots,7\}$ and $R_i(0) \in \{3,\ldots,6\}$, $\forall i$, respectively. 
The measured final MSE values~\eqref{eq:se-mse} for these schemes were set as 
the target values for the GGP.

For all schemes, $\MSE{\Dv, T}$ and $\Ragg$ were computed. These values were 
averaged over all 32 realizations of each ($\rho_c$, $R_i(0)$, $T$) setting, and 
the resulting averages were plotted against each other.  
Looking at \eqref{eq:se-data-simple}, it 
seems that the MSE for
quantized consensus, where $D_i(t)\geq 0, \forall i,t$, is greater than in unquantized consensus, where $D_i(t) = 0, \forall i,t$.
 
We therefore introduce two terms to define the MSE performance relative 
to the ideal, unquantized algorithm. To compensate for the effect of network 
topology on the MSE, we define the {\em lossless MSE},
\begin{align}
\text{MSE}_{\text{lossless}}(t) := \MSE{\Dv,t} \Big\vert_{\Dv=\mathbf{0}}.
\end{align}
Next, define the final {\em excess MSE} (EMSE) as
\begin{align}
\text{EMSE}(T) := 10 \log_{10} \frac{\text{MSE}(\Dv,T)}{\text{MSE}_{\text{lossless}}(T)},
\end{align}
which represents the increase in MSE over lossless consensus resulting from 
distortion. The EMSE was used for the generation of RD trade-off curves.

In the case of very low rates using ECSQ on zero-mean Gaussian
sources,  
all elements decoded at a receiving node would be zeros. To 
prevent this behavior, the maximum normalized distortion $D_{\text{max}}$ 
allowed was set such that the recieved elements were nonzero at least 1\% 
of the time.

Because of stability issues with the GGP solvers used, Yildiz and Scaglione's 
order-one predictive coding implementation~\cite{YildizScaglione2008}, which was 
provided by the authors, was modified to use fixed-rate uniform 
quantization but allow for the rate to vary with iteration and node indices. 
This capability was implemented by running two rate update recursions---one to 
keep track of the ideal (real-valued) rates given by the quantization noise 
variance recursion~\cite{YildizScaglione2008}, and another to perform the 
predictive coding using rates that were rounded to the nearest integral value.

\section{Discussion}


To adequately discuss the RD results, we first comment on some properties of 
each prior art scheme presented. The ProgQ algorithm~\cite{Thanou2013} uses a 
time- and node-invariant fixed-rate uniform quantizer (i.e., $R_i(t) = R$, 
$\forall i \in \{1,\ldots,m\}, t \in \{0, \ldots, T-1\}$), whereas Yildiz and 
Scaglione~\cite{YildizScaglione2008} allow the use of different rates at each 
node and iteration. 

Thanou \etal~\cite{Thanou2013} use the same state update as 
ours~\eqref{eq:lossy-iteration}, but Yildiz and 
Scaglione~\cite{YildizScaglione2008} use a different update that is 
incapable of truly converging in the presence of quantization error. The 
final asymptotic MSE for the predictive scheme depends on the sum of 
distortions, $D_i(t), t \in \mathbb{Z}_{\geq 0}$. If these distortions are 
chosen to form a convergent series, then the MSE will converge to a nonzero, 
but bounded, value. Because of this limitation, the predictive 
scheme~\cite{YildizScaglione2008} is heavily dependent on the starting rates, 
$R_i(0)$.

In some cases, such as the bottom right of the ECSQ curve in 
Fig.~\ref{fig:tradeoff-curve-ysp1}, the predicted performance and measured 
performance of ECSQ do not match. Because the ECSQ used in the simulations 
is not dithered~\cite{Lipshitz1992}, the additive quantization model only holds 
approximately. As $\Ragg$ increases, the performance improves, and better 
adherence to predicted performance can be accomplished using 
dithering~\cite{Lipshitz1992}.

The RD performance of the proposed optimization scheme~\eqref{eq:cfun-mod} 
for ECSQ is 
compared to the predictive coding scheme of Yildiz and 
Scaglione~\cite{YildizScaglione2008} and the ProgQ algorithm of 
Thanou \etal~\cite{Thanou2013} in Fig.~\ref{fig:tradeoff-curve-ysp1}.  For 
our scheme, the predicted RD performance (as computed by the state evolution 
equations~\cref{\seeqs}) is compared to the actual performance. The distortion 
(measured by the EMSE) is given by the horizontal axis, and the aggregate rate 
$R_{\text{agg}}$ by the vertical axis. The predicted performance is denoted by 
a dashed line, while the simulated performance is represented as a solid line. 
Alongside our approaches, we plot the RD performance of both of the comparators. 
A curve closer to the bottom left corner of these figures indicates better 
performance, meaning lower aggregate rate $R_{\text{agg}}$ to achieve the same 
EMSE, or lower EMSE for a particular $R_{\text{agg}}$.

The numerical results in the left panel of Fig.~\ref{fig:tradeoff-curve-ysp1} 
suggest that our GGP approach outperforms that of 
Yildiz and Scaglione and Thanou \etal~However, a closer look reveals that much 
if not all of the gain is due to our using variable rate coding, whereas the 
implementations for the comparators use fixed rate coding. When we evaluated our 
approach with fixed rate coding using a heuristic proposed in
Pilgrim~\cite{PilgrimThesis2017}, our results were typically somewhat weaker 
than the comparators. We attribute the performance advantage of 
ProgQ~\cite{Thanou2013} and predictive coding~\cite{YildizScaglione2008} to
their use of side information from previous iterations. 

For a fairer comparison to the predictive approach of Yildiz and 
Scaglione~\cite{YildizScaglione2008}, we also simulated their approach against 
ours~\eqref{eq:cfun-mod} for RD-optimal vector 
quantizers, which use variable coding rates. The 
simulations were 
run on a single instance of the RGG by varying the initial coding rate of the 
predictive scheme~\cite{YildizScaglione2008} and setting the resulting 
$\text{EMSE}(T)$ as the target for the optimization~\eqref{eq:cfun-mod}. Because 
the quantization error of infinite-dimensional lattice quantizers approaches an 
additive white Gaussian noise process~\cite{ZamirFeder96}, these experiments 
simulated quantization by adding independent noise of variance $D(t)$ to the 
states. This plot 
demonstrates the advantage of predictive quantization on an even playing 
field by allowing both our approach and the predictive 
scheme~\cite{YildizScaglione2008} to use variable-rate coding.

The ProgQ scheme~\cite{Thanou2013}, unlike predictive 
coding~\cite{YildizScaglione2008}, is capable of converging in the limit due to 
its different state update strategy. Both ProgQ~\cite{Thanou2013} and 
predictive coding~\cite{YildizScaglione2008} are capable of using 
constant or even shrinking coding rates to achieve good performance. It is 
clear from Fig.~\ref{fig:varD-optimal-rates} that our rates grow with $t$. 
Therefore, as $T \rightarrow \infty$, we expect that ProgQ will outdo our 
proposed schemes in all settings, despite our constrained optimization, because
it can converge in the limit of large $T$ with constant rates.

\section{Conclusion}
In conclusion, this paper presented a framework for optimizing the source coding 
performance of 
distributed average consensus. The key 
insight of our approach is the formulation of the problem as a 
GGP~\cite{BoydKimVandenbergheHassibi2007}. Our framework allows the problem to 
be transformed to a convex program~\cite{BoydVandenberghe2004} and solved for 
the global optimum. Although we do not incorporate knowledge from
past iterations, our numerical results are competitive with prior art that
uses more 
sophisticated side information strategies, which motivates the study of
optimization for
predictive coding schemes.

In light of the performance gain from predictive coding strategies, 
we feel that future work should focus on variable rate strategies. Moreover, 
we aim to optimize the predictive approach of Yildiz and Scaglione using our 
GGP formulation and the state update~\eqref{eq:lossy-vector-iteration}. 

\section*{Acknowledgments}
Thanks to Yanting Ma for her inputs on extending the GGP model to variable
distortion, to Mehmet Ercan Yildiz and Anna Scaglione for graciously 
providing their code for comparison, and to Yaoqing Yang and Pulkit Grover for 
discussing their work with Soummya Kar.

\IEEEtriggeratref{17}
\bibliography{corrected3}

\end{document}